\documentclass[%
reprint,
amsmath,amssymb,
aps,
pra,
]{revtex4-2}

\usepackage{graphicx}
\usepackage{dcolumn}
\usepackage{bm}
\usepackage{empheq} 
\usepackage{fancybox} 

\usepackage[english]{babel}
\makeatletter
\let\ORIbbl@fixname\bbl@fixname
\def\bbl@fixname#1{%
	\@ifundefined{languagealias@\expandafter\string#1}
	{\ORIbbl@fixname#1}
	{\edef\languagename{\@nameuse{languagealias@#1}}}%
}
\newcommand{\definelanguagealias}[2]{%
	\@namedef{languagealias@#1}{#2}%
}
\makeatother
\definelanguagealias{en}{english}
\definelanguagealias{EN}{english}
\definelanguagealias{eng}{english}
\definelanguagealias{fr}{english}

\begin{document}


\title{Frequency-modulated combs as phase solitons}

\author{David Burghoff}
\affiliation{%
 Department of Electrical Engineering, University of Notre Dame, Notre Dame, IN 46656\\
}%

%
%

\date{\today}

\begin{abstract}
Frequency combs are light sources with coherent evenly-spaced lines. It has been observed that in certain laser systems, combs can form whose output is frequency-modulated (FM) in time. In this state, they produce an output whose frequency sweeps linearly and periodically. These results have been replicated numerically, but a thorough understanding of their core physics remains elusive. Surprisingly, we have found that these lasers are described by a nonlinear Schr\"{o}dinger equation whose potential is proportional to the phase of the electric field. This equation can be solved exactly and produces a field whose phase is piecewise quadratic in time---an FM comb. These results can be used to derive all of the salient features of FM combs, and our general theory is applicable to any nonlinear optical system with large internal gain. More generally, this result portends the development of new coherent states of light governed by phase potentials rather than amplitude potentials.

\end{abstract}

\pacs{Valid PACS appear here}
\maketitle


\section{\label{sec:level1}Introduction}

Frequency-modulated (FM) combs are a type of frequency comb that have been produced in a number of laser systems. First observed in the 1960s in electro-optically modulated cavities \cite{harrisFMOSCILLATIONHe1964}, such lasers were known to not produce pulses but would instead produce an FM output \cite{yarivInternalModulationMultimode1965}. These systems were assumed to have sinusoidal modulation, but no direct measurements of the temporal output could be performed since the peak intensities were too low. More recently, it has been shown that many lasers will spontaneously enter self-FM regimes, where the FM is produced without any external modulation and the amplitude is approximately constant. Initially observed in quantum cascade lasers (QCLs)s \cite{hugiMidinfraredFrequencyComb2012,khurginCoherentFrequencyCombs2014,burghoffTerahertzLaserFrequency2014,singletonEvidenceLinearChirp2018}, this mode of operation has since been extended to other types of lasers, including quantum dot lasers \cite{hillbrandInPhaseAntiPhaseSynchronization2020} and diode lasers \cite{sterczewskiFrequencymodulatedDiodeLaser2020}. While not pulses, they can still be used in applications like comb spectroscopy \cite{villaresDualcombSpectroscopyBased2014,yangTerahertzMultiheterodyneSpectroscopy2016,burghoffComputationalMultiheterodyneSpectroscopy2016}.

Earlier FM comb observations relied only on observations of a narrow coherent beatnote and of a broadband spectrum. However, it is well known that these observations are not sufficient to fully reconstruct the temporal profile. Indeed, early reports of pulse formation in QCLs \cite{paiellaSelfModeLockingQuantumCascade2000} were later determined to have been caused by coherent instabilities \cite{gordonMultimodeRegimesQuantum2008}. It was only recently, with the development of Shifted Wave Interference Fourier Transform Spectroscopy (SWIFTS) \cite{burghoffBroadbandTerahertzPhotonics2014,burghoffEvaluatingCoherenceTimedomain2015,hanSensitivitySWIFTSpectroscopy2020}, that it became possible to measure high-quality temporal traces of low-intensity combs. Surprisingly, it was found that many of these FM combs do not have sinusoidal modulation, they have linear-chirped operation \cite{singletonEvidenceLinearChirp2018,hillbrandInPhaseAntiPhaseSynchronization2020,sterczewskiFrequencymodulatedDiodeLaser2020,hanSensitivitySWIFTSpectroscopy2020} (or boxcar operation when the gain spectrum is discontinuous \cite{burghoffEvaluatingCoherenceTimedomain2015}). This chirp is completely coherent and comprises the whole laser spectrum.

While linear-chirped behavior is a robust result that has been replicated in many different systems, it is not fully understood. Primarily, these systems are simulated using Maxwell-Bloch formalisms, which lead to a series of coupled partial differential equations \cite{gordonCoherentTransportSemiconductor2009,wangActiveModelockingMidinfrared2015,silvestriCoherentMultimodeDynamics2020,wangHarmonicFrequencyCombs2020,tzenovAnalysisOperatingRegimes2017,dongPhysicsFrequencymodulatedComb2018}. Alternatively, they can be simulated using modal expansions of optical nonlinearities \cite{khurginCoherentFrequencyCombs2014,henryPseudorandomDynamicsFrequency2017,henryTemporalCharacteristicsQuantum2018}. While these equations are fully descriptive and can capture FM operation, they are often difficult to understand. More recently, Opačak and Schwarz \cite{opacakTheoryFrequencyModulatedCombs2019} showed that a master equation description can capture the physics of FM comb formation, impressively reducing the number of coupled equations from eight to two. These results were numerical and were able to produce linear FM operation as well as more complex behavior \cite{piccardoFrequencyCombsInduced2020}, but do not show precisely why linear operation occurs. In addition, these results require the integration of many small time steps, requiring long simulation times. Gaining a fundamental of this behavior is critical for improving the performance of FM combs beyond dispersion engineering \cite{burghoffDispersionDynamicsQuantum2016}.

In this work, we derive a mean-field theory that fully describes FM operation and even admits exact solutions under certain conditions. This theory is analogous to the celebrated Lugiato-Lefever Equation (LLE) used to describe many nonlinear resonators, such as microresonator combs \cite{kippenbergDissipativeKerrSolitons2018,yangBroadbandDispersionengineeredMicroresonator2016,delhayeOpticalFrequencyComb2007,okawachiBandwidthShapingMicroresonatorbased2014} and other Kerr combs \cite{yuPhotonicCrystalReflectorNanoresonatorsKerrFrequency2019}. This theory uses the concept of the extended-cavity theory that has been used to describe the dynamics of nonlinear Fabry-Perot cavities \cite{lugiatoTravelingWaveFormalism2018,coleTheoryKerrFrequency2018}, but with one distinct difference---it takes into account the large changes of the field that are present in a laser cavity with large mirror losses and spatial hole burning. The resulting equation is integral and allows for a single time step to be taken per round trip. Importantly, we show that near equilibrium, the electric field of the system can be described by a nonlinear Schr\"{o}dinger equation (NLSE) with a potential proportional to its \textit{phase}:

\begin{align} \label{nlse}
	-i \frac{\partial E}{\partial t} = \frac{\beta}{2} \frac{\partial^2 E}{\partial z^2} + \gamma |E|^2 (\arg E-\langle \arg E \rangle) E \nonumber \\ +i r (\left|E\right|^2-P_0),
\end{align}
where $\beta$ is the normalized dispersion, $\gamma$ is the nonlinear cross-steepening, and r represents amplitude relaxation. Unlike the conventional nonlinear Schr\"{o}dinger equation, whose potential depends on the intensity of the pulse, here it is primarily determined by its phase. This equation can be solved analytically as
\begin{align} \label{solitoneqn}
	E(z,t) =A_0 \exp\left[i \frac{\gamma |A_0|^2}{2 \beta} \left(z^2-\frac{1}{3}L_c^2 \gamma |A_0|^2 t\right) \right]
\end{align}
where $A_0=\sqrt{P_0}(1+\frac{\gamma}{2 r})^{-1/2}\approx \sqrt{P_0}$ is the soliton's amplitude, $L_c$ is the physical cavity length, and $z\in(0,2L_c)$ is the position within the artificially extended cavity. Equation \ref{solitoneqn} is the fundamental soliton of FM combs. Like the celebrated sech$^2$ solitons that result from an intensity potential, this soliton is nonperturbative and can be used as a starting point for more sophisticated analysis. As it results from phase rather than amplitude, we refer to it as a phase soliton.

We will show that this result can be used to understand many of the salient features of FM combs. For example, this result explains the pulsation that is frequently observed at the turnaround point where the frequently abruptly changes. It explains why these structures are not observed in other nonlinear media with self-steepening, such as microresonators and fibers. It also allows us to analytically examine the conditions under which FM combs can form.

This work proceeds as follows. In Section \ref{mftheory}, a mean field theory analogous to the LLE is derived for lasers with large intracavity dynamics. The essential idea is to replace this large variation with a slowly-varying envelope, which is then integrated over a round trip. In Section \ref{secnlse}, we show how this equation can be simplified with some weak assumptions on the steady-state intensity of the laser, giving rise to equation \ref{nlse}. Finally, in Section \ref{sec:numerical} we perform numerical simulations of the mean field theory, comparing the result to the analytical results. We also discuss the conditions under which these combs are stable.

\section{Mean field theory} \label{mftheory}

First, we derive a mean-field equation that describes the propagation of light within lasers with large spatial hole burning and mirror losses. Mean-field theory is a powerful technique that allows for the dynamics of nonlinear resonators to be evaluated after a single round trip through the resonator, replacing the much smaller timesteps that would otherwise be required. Our starting point are the master equations for a semiconductor laser with cross-steepening:
\begin{figure}
	\includegraphics[width=\linewidth]{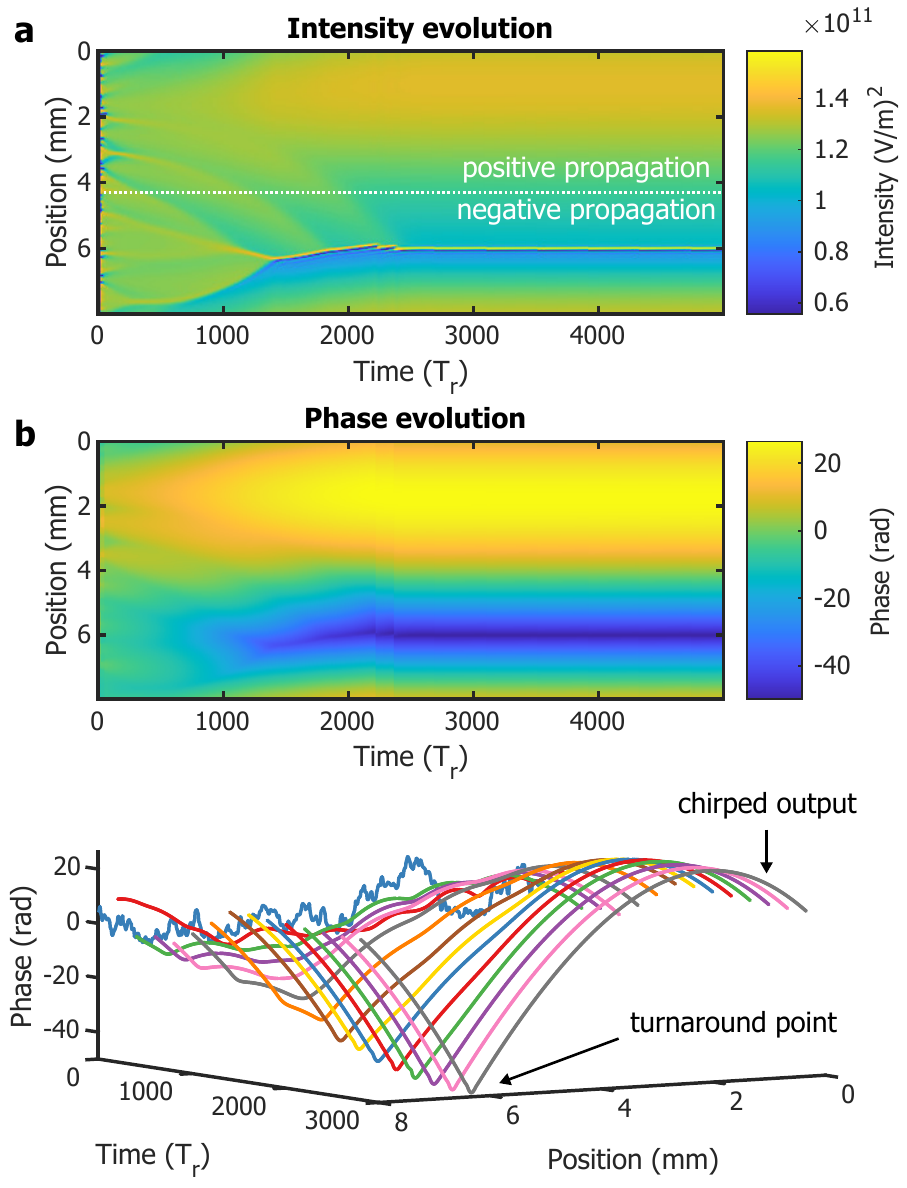}
	\caption{\label{fig:splitstep}Split step simulation of a QCL using the mean-field theory described by equation (\ref{mylle}). Parameters are given in Table \ref{tab:params}, and QCL has a GVD of -2000 fs$^2$/mm and zero Kerr nonlinearity. Position values below $L_c=$4 mm represent positive propagation, values above $L_c$ represent negative propagation. a. Evolution of the intracavity intensity over 5000 round trips. The intensity quickly builds to a constant value from spontanenous emission, then gains some modulation due to comb formation. b. Evolution of the phase. The phase begins random, and eventually gives way to a periodic parabola with an abrupt turnaround point.}
\end{figure}
\small
\begin{align} \label{fulleqn} 
	\frac{n}{c} \frac{\partial E_\pm}{\partial t} + \frac{\partial E_\pm}{\partial z} = \frac{g_0}{2}\left(1-\frac{1}{P_s}(\left|E_\pm\right|^2+2\left|E_\mp\right|^2) \right) E_\pm \nonumber \\ -\frac{\alpha_w}{2} E_\pm +i\frac{1}{2}k''\frac{\partial^2 E_\pm}{\partial t^2} + \frac{g_0}{2}T_2^2 \frac{\partial^2 E_\pm}{\partial t^2}\nonumber \\
	 - i \gamma_K (\left|E_\pm\right|^2+2\left|E_\mp\right|^2)E_\pm   \\ 
	+\frac{g_0}{2 P_s}\left((2T_1+3T_2) \frac{\partial E_\mp^\ast}{\partial t} E_\mp +(T_1+\tfrac{5}{2}T_2) E_\mp^\ast \frac{\partial E_\mp}{\partial t} \right)E_\pm \nonumber
\end{align}
\normalsize where $E_\pm$ is the envelope of the wave traveling in the $\pm$z direction, $g_0$ is the small signal gain, $k''$ is the group velocity dispersion (GVD), $\gamma_K$ is the Kerr nonlinearity, and $T_1$ and $T_2$ are the population and coherence lifetimes within the Bloch equations for a two-level system. This result is similar to the one obtained in Ref. \cite{opacakTheoryFrequencyModulatedCombs2019}, but is instead derived from an expansion of the solution to the Bloch equation, expanded to first order in $\frac{\partial}{\partial t}$ and third-order in $E_\pm$. In addition, the $\frac{g_0}{2}T_2^2\frac{\partial^2 E_\pm}{\partial t^2}$ is kept whenever the effects of gain curvature are considered. Several unimportant terms are neglected for now; see Appendix \ref{sec:meqn} for the full version. The main terms that are critical for FM comb formation are the dispersive term and the terms in the last line, which we refer to as \textit{cross-steepening terms}. Like the self-steepening that occurs in fibers \cite{agrawalNonlinearFiberOptics2012}, they can be construed as a Raman effect arising from an intensity-dependent group delay, i.e., $\frac{\partial}{\partial t} |E_\mp|^2 E_\pm$. However, cross-steepening instead arises from the intensity of the counterpropagating wave.

Equation (\ref{fulleqn}) is general and can be numerically integrated, but is difficult to analyze within a mean-field formalism. The challenge is that the changes that occur in lasers within a round trip are large and cannot be neglected, unlike in other nonlinear resonators. Likewise, the mirror losses cannot be neglected, as is commonly done \cite{lugiatoLugiatoLefeverEquation2018,coleTheoryKerrFrequency2018}. To account for this, we make several additional modifications:
\begin{enumerate}
	\item Backward propagating waves are flipped and extended to $z\in(L_c,2 L_c)$. This reduces the problem to a single forward-propagating wave, $E(z,t)$, that is periodic with period $2L_c$.
	\item The steady-state intensity $P(z)$ is found only in the presence of gain saturation, waveguide loss, and mirror losses. The effective gain is defined as $g_{\textrm{eff}}(z) \equiv -\alpha_w + g_0\left(1-\frac{1}{P_s}(P(z)+2P(-z) \right) + \ln(R_1) \delta(z)+\ln(R_2) \delta(z-L_c)$. Let $P_0=P(0)$.
	\item The slow-intensity envelope $F(z,t)$ is defined in terms of the effective gain as $E(z,t) \equiv F(z,t) \exp \frac{1}{2} \int_{0}^{z} g_{\textrm{eff}}(z') dz'\equiv F(z,t) K^{1/2}(z)$. Unlike the bare electric field, the slow-intensity envelope is approximately constant (even at the mirrors, where $E(z,t)$ is discontinuous). 
\end{enumerate}

$K(z)$ can be understood as the dimensionless power gain that a field experiences when propagating from 0 to z, i.e. $P(z)=P_0 K(z)$. Making these substitutions, we find that the slow-intensity envelope F(z,t) cancels the loss and gain saturation terms, leaving only
\small
\begin{align} \label{fbeqn} 
	\frac{n}{c} \frac{\partial F}{\partial t} + \frac{\partial F}{\partial z} =  
	-\frac{g_0}{2P_s}\left(K \left|F\right|^2-P+2K_-\left|F_-\right|^2 - 2 P_{-} \right) F \nonumber \\
	+i\frac{1}{2}k''\frac{\partial^2 F}{\partial t^2} + \frac{g_0}{2}T_2^2 \frac{\partial^2 F}{\partial t^2} - i \gamma_K \left(K \left|F\right|^2+2K_-\left|F_-\right|^2 \right)F\nonumber \\
	+\frac{g_0}{2 P_s}K_-\left((2T_1+3T_2) \frac{\partial F_-^\ast}{\partial t} F_- +(T_1+\tfrac{5}{2}T_2) F_-^\ast \frac{\partial F_-}{\partial t} \right)F.
\end{align}
\normalsize where we have used the notation that minus signs indicate negative position, e.g. $K_- \equiv K(-z)$. At this point we have made no approximations; this equation is identical to (\ref{fulleqn}). Because we buried all steady-state losses into K, gain now looks like a mere restoring force. From a numerical perspective, this equation is already simpler, as it is a single equation with periodic boundary conditions.

To develop mean-field equations analogous to the LLE, we must integrate (\ref{fbeqn}) over a single round trip through the cavity. This is the first approximation that is made, as we are implicitly using the slowly-varying nature of F. While this integration is simple for terms in the positive coordinate alone, any term with a backward coordinate will give rise to a convolution because within the extended cavity formalism it depends non-locally on the electric field. Defining $\langle K \rangle \equiv \frac{1}{2L_c}\int_{0}^{2 L_c}K(u) du$ and the convolution $\tilde{K}[f](z)\equiv\frac{1}{4L_c}\int_{0}^{4 L_c}K(-\frac{u}{2})f(z-u) du$, we then find that the change over a round trip is given by
\newcommand*\widefbox[1]{\fbox{\hspace{0.5em}#1\hspace{0.5em}}}
\begin{align} \label{mylle} 
	\frac{\partial F}{\partial t} = 
	&-\frac{1}{3}r\left(\left|F\right|^2 + 2\langle K \rangle^{-1}\tilde{K}[|F|^2] - 3 P_0 \right) F \nonumber \\
	&+i\frac{1}{2}\beta\frac{\partial^2 F}{\partial z^2} + D_g \frac{\partial^2 F}{\partial z^2} \nonumber \\
	&- i \frac{1}{3}\gamma_K' \left(\left|F\right|^2 + 2\langle K \rangle^{-1}\tilde{K}[|F|^2] \right)F\nonumber \\
	&-\frac{g_0}{2 P_s}\left(\frac{c}{n}\right)^2\tilde{K}\bigg[(2T_1+3T_2) \frac{\partial F^\ast}{\partial z} F \nonumber\\ &\hspace{2.2cm}+(T_1+\tfrac{5}{2}T_2)F^\ast \frac{\partial F}{\partial z} \bigg]F,
\end{align}
where $\beta\equiv k''\left(\frac{c}{n}\right)^3$ is the normalized dispersion, $r\equiv3\frac{g_0}{2 P_s}\frac{c}{n} \langle K \rangle$ represents energy relaxation, $D_g \equiv \frac{g_0}{2}T_2^2\left(\frac{c}{n}\right)^3$ is the gain curvature, and $\gamma_K'\equiv  3 \gamma_K \frac{c}{n} \langle K \rangle$ is a normalized Kerr nonlinearity. Note that on the right hand side we have also made a spatiotemporal substitution \cite{chemboSpatiotemporalLugiatoLefeverFormalism2013}, replacing fast time derivatives with spatial derivatives ($\partial_t \rightarrow -\frac{c}{n} \partial_z$). This is our final result.

In comparison with the standard LLE and the FP-LLE formalisms derived for propagation within low-loss cavities \cite{coleTheoryKerrFrequency2018,lugiatoLugiatoLefeverEquation2018,lugiatoSelfpulsingFabryPerotLasers2019}, we note that our formalism is somewhat more complicated due to the presence of K(z). However, this is essential for understanding FM comb operation, as it is the spatial dependence of K(z) that gives rise to stable FM comb formation. Though this equation is integral in nature and is difficult to analyze as written, it is highly amenable to numerical analysis since $\tilde{K}\left[f\right]$ is a periodic convolution and can be rapidly evaluated using Fourier methods. It can be readily solved using split-step methods, as is commonly done in fibers and other nonlinear resonators.

An example of simulating a typical mid-infrared QCL comb with a split-step implementation of the mean-field theory is shown in Figure \ref{fig:splitstep}.
The field is initialized to a low value and initially stabilizes to its steady-state intensity, but over many round trips the phase builds and eventually gives way to a periodic parabola, representing an FM comb (whose frequency is linearly-chirped). Because we need only take one time step per round trip, the simulation takes only seconds.

\section{Nonlinear Schr\"{o}dinger equation with a phase potential} \label{secnlse}

Next, we show how our mean field theory leads to a nonlinear Schr\"{o}dinger equation with a phase potential. To do this, we ignore the Kerr nonlinearity and gain curvature terms. Without gain curvature, the amplitude is essentially constant in space everywhere but the turnaround point. Let A and $\phi$ be F's amplitude and phase. The derivatives in the cross-steepening term can be written as $\frac{\partial F^\ast}{\partial z}F=-i \frac{\partial \phi}{\partial z}|A|^2$, and the nonlinear gain simply becomes
\begin{align}
g_\textrm{NL} &= i\frac{g_0}{2 P_s}\frac{c}{n}(T_1+\frac{1}{2}T_2)|A|^2\tilde{K}\left[\frac{\partial \phi}{\partial z}\right] \nonumber\\
&= i\frac{g_0}{2 P_s}\frac{c}{n}(T_1+\frac{1}{2}T_2)|A|^2 \nonumber \\
&\times\frac{-1}{4L_c P_0}\int_{0}^{4 L_c}\frac{\partial P}{\partial z}\left(-\frac{u}{2}\right)\phi(z-u)du
\end{align}
\begin{figure*}
	\includegraphics[width=0.95\linewidth]{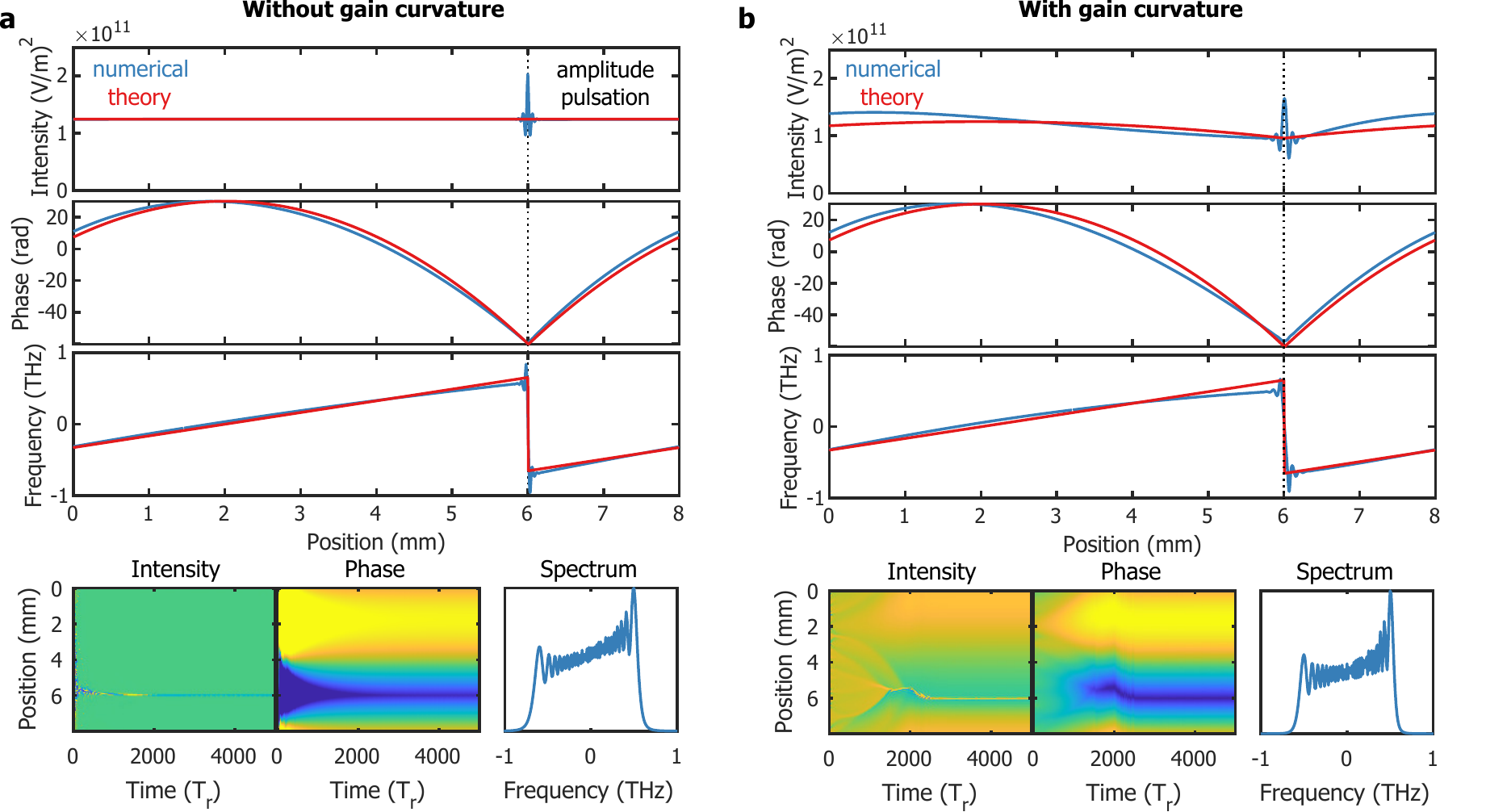}
	\caption{\label{fig:analyticcomp}a. Comparison of mean-field simulation results and soliton theory with gain curvature disabled ($k''=$-1700 fs$^2$/mm and $\gamma_K=0$). The agreement between the simulation and soliton theory is excellent, correctly predicting the chirp. The theory breaks down at the turnaround point, where an amplitude pulsation develops. b. Same comparison, but with gain curvature. As the FM deviates from it center frequency, the amplitude sags due to the lower gain (and can be estimated by equation (\ref{ampsag})).}
\end{figure*}
Proceeding further requires some knowledge of the steady-state power profile. For an asymmetric cavity with $R_2$ = 1, we approximate the power profile as piecewise linear. The most important feature of this curve is \textit{not} the shape of the curve itself, but the discontinuity that occurs when the power reflects at mirror 1. Practically identical results are obtained even when the curve bows or when the mirror losses are split over two mirrors rather than one. If the change in the power at mirror 1 is denoted by $\Delta P$, then
\begin{align}
\frac{\partial P}{\partial z} &= \frac{\Delta P}{2 L_c}-\Delta P \delta(z) \nonumber\\
g_\textrm{NL} &= i\frac{g_0}{2 P_s}\left(\frac{c}{n}\right)^2(T_1+\frac{1}{2}T_2)\frac{|F|^2}{P_0} \frac{\Delta P}{4L_c}(\phi-\langle\phi\rangle)
\end{align}
After convolution, the linear part of the power profile led to the average value of the phase, while the discontinuity at the mirror led to its instantaneous value. Thus, the discontinuity can be considered the key ingredient for FM comb formation. After a wave has reflected from the facet, it sees a cross-steepening term that represents its phase at a slightly earlier time. This induces a modulation instability that causes the phase to undergo positive feedback. Similarly, the amplitude relaxation term can be integrated to produce the final NLSE:
\begin{align}
-i\frac{\partial F}{\partial t} &= \frac{\beta}{2} \frac{\partial^2 F}{\partial z^2}+\gamma|F|^2(\phi-\langle\phi\rangle)F+ i r(|F|^2-P_0)F
\end{align}
where $\gamma\equiv\frac{g_0}{2 P_s}\left(\frac{c}{n}\right)^2(T_1+\frac{1}{2}T_2) \frac{\Delta P}{4L_c P_0}$ is the normalized cross-steepening nonlinearity. By solving for the amplitude and phase separately, one can show that an analytical solution is
\begin{align} \label{solitoneqn2}
E(z,t) =A_0 \exp\left[i \frac{\gamma |A_0|^2}{2\beta} \left(z^2-\frac{1}{3}L_c^2 \gamma |A_0|^2 t\right) \right]
\end{align}
where $A_0=\sqrt{P_0}(1+\frac{\gamma}{2 r})^{-1/2}\approx \sqrt{P_0}$. This is the key result, as this solution is nonperturbative and almost completely describes FM combs. At also describes higher order Turing rolls (a.k.a. harmonic states \cite{mansuripurSinglemodeInstabilityStandingwave2016}). Its instantaneous frequency can be read off directly as 
\begin{align} \label{freqs}
f_i(z) &= -\frac{c}{2\pi n}\frac{\gamma}{\beta}|A_0|^2 z  \\
f_i(t) &= f_i(-\frac{c}{n}t) = \frac{1}{2\pi}\left(\frac{c}{n}\right)^2\frac{\gamma}{\beta}|A_0|^2 t  \\
\end{align}
In other words, the laser is chirped linearly in time by an amount inversely proportional to its dispersion. This simple formalism predicts that zero dispersion produces a comb of \textit{infinite bandwidth}, but as we will show later a finite gain curvature will not allow a stable comb to exist unless the sweep bandwidth is limited. We can also use this formalism to compute the carrier envelope offset (CEO) of the comb. The time-dependent term in (\ref{solitoneqn2}) gives rise to a cross-steepening contribution to the CEO, and the Kerr nonlinearity adds another (calculated using the mean field theory):
\begin{align} 
f_\textrm{CEO,XS}&=-\frac{\gamma^2 |A_0|^4}{12 \pi \beta} L_c^2 \\
f_\textrm{CEO,Kerr} &= -\frac{1}{2\pi}\gamma_K \frac{c}{n}\frac{3\langle P \rangle}{P_0}|A_0|^2 \label{kerrceo}
\end{align}
For typical mid-IR QCLs, with values of $\gamma_K$ on the order of $10^{-10}$ m/V$^2$, the cross-steepening induced shift is on the order of tens of GHz, while the Kerr-induced shift is on the orders of GHz.

\section{Results and discussion} \label{sec:numerical}

First, we discuss why the phase ends up linearly-chirped. Consider the time evolution of the phase component to the NLSE:
\begin{align} \label{phaseevolve}
\frac{\partial \phi}{\partial t}=-\frac{\beta}{2}\left(\frac{\partial \phi}{\partial z}\right)^2 + \gamma |A|^2 (\phi-\langle\phi\rangle)
\end{align}
Initially, the dispersive term is small, and phase perturbations experience exponential gain due to the nonlinearity. If the GVD is negative, negative perturbations are suppressed when the perturbations grow sufficiently large, and positive perturbations are enhanced. However, at the FM turnaround point the derivative is zero, and the exponential gain in the negative direction remains uninhibited. This generates a discontinuity in the derivative of $\phi$, which leads to the production of a characteristic pulsation in the amplitude. This pulsation is physical and has been observed in several systems \cite{singletonEvidenceLinearChirp2018,sterczewskiFrequencymodulatedDiodeLaser2020}, but can ultimately destabilize the soliton.

Figure \ref{fig:analyticcomp} compares the analytical results of the phase soliton solutions to the solutions obtained by the full mean-field theory. First, we consider the case without gain curvature, shown in Figure \ref{fig:analyticcomp}a.
The agreement between the mean-field theory and the analytic form is excellent, as the assumption of a linear intracavity power is actually a relatively weak one. The most conspicuous discrepancy is present at the turnaround point, where the assumption of a constant amplitude cannot hold, as the second derivative of $\phi$ is not well-defined. Provided the pulsation is small, the soliton will be able to form, as spatial inhomogeneities are automatically dampened by the relaxation term. However, consider the dispersive term, whose phase evolution is given by $\frac{\partial \phi}{\partial t}=\frac{\beta}{2}\big(\frac{1}{A} \frac{\partial^2 A}{\partial z^2}-\big(\frac{\partial \phi}{\partial z}\big)^2\big)$ when inhomogeneities are present in the amplitude. If the pulsation causes the amplitude to dip to near zero, amplitude inhomogeneities are magnified and begin to destabilize the phase. When these fluctuations become sufficiently large, they will become self-reinforcing and break comb operation.

\begin{figure*}
	\includegraphics[width=0.95\linewidth]{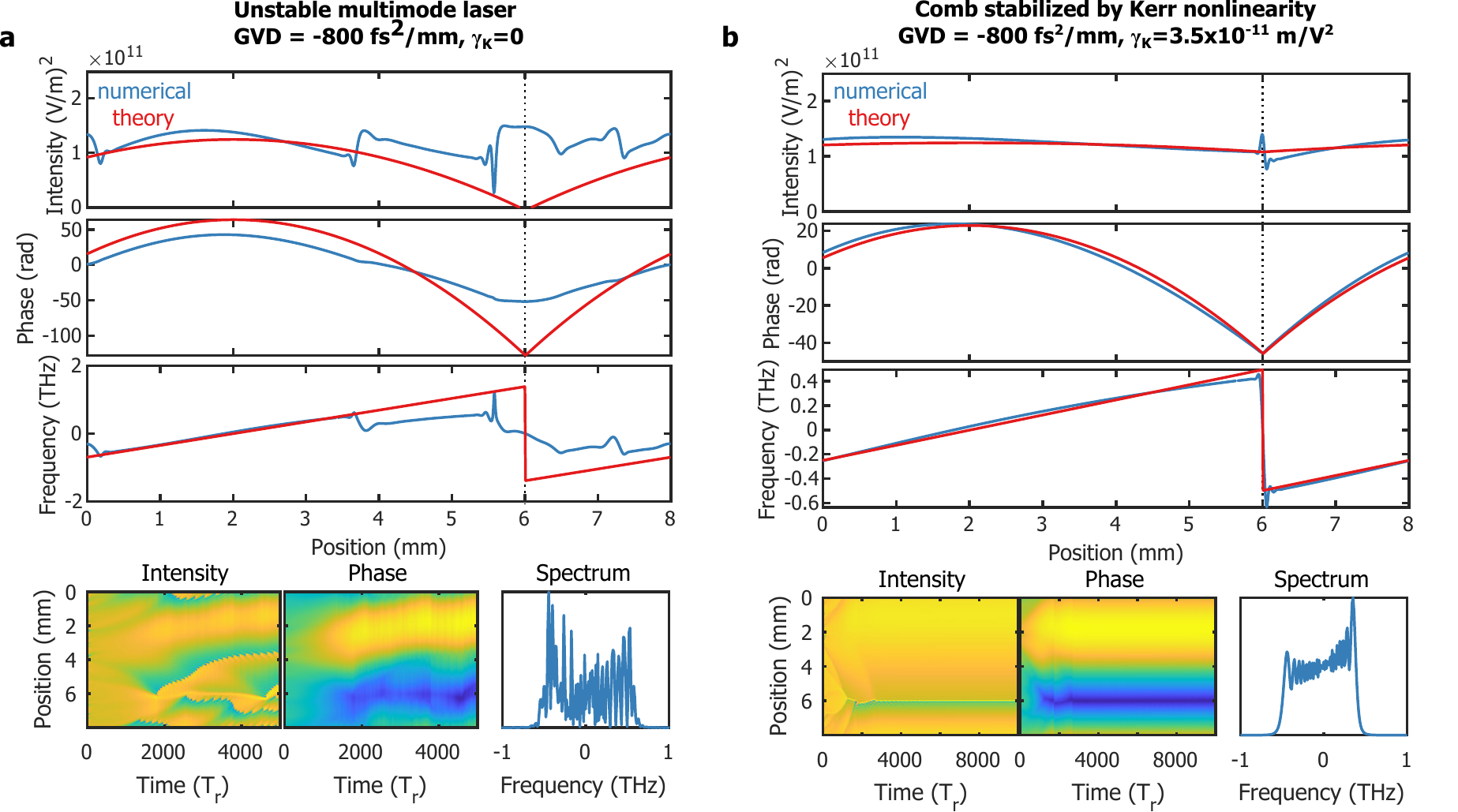}
	\caption{\label{fig:kerrcurvature} a. Simulation and theory for a laser with zero Kerr nonlinearity and a dispersion that is too low to sustain comb operation. A stable comb cannot form, because as the theory predicts the intracavity intensity would hit zero near the turnaround point. b. Corresponding laser with a Kerr nonlinearity of 3.5$\times10^{-11}$ m/V$^2$. The Kerr effect reduces the effective dispersion of the laser according to (\ref{deff}), leading to an effective dispersion of -2230 fs$^2$/mm. A stable comb can now form.}
\end{figure*}
In the absence of gain curvature, there is essentially no value of the dispersion that cause the comb to destabilize---the NLSE supports infinite-bandwidth FM combs. However, once gain curvature is considered, this is no longer the case. Fig. \ref{fig:analyticcomp}b shows the same results as in \ref{fig:analyticcomp}a, but with gain curvature enabled. Characteristic dips in the amplitude are now visible around the pulsation, which occur because the FM comb has deviated from its central frequency and the gain has been reduced. Thus, the power begins to fall. If the dispersion is reduced such that the FM comb bandwidth exceeds the gain bandwidth of the laser, the power will dip to near zero and the amplitude pulsation will destabilize comb operation. The change in power induced by gain curvature can be estimated in equilibrium as
\begin{align} \label{ampsag}
\delta P = -\frac{D_g}{r} \left(\frac{\partial \phi}{\partial z}\right)^2,
\end{align}
leading to the following condition for stable soliton formation:
\begin{align}
P_0 \frac{D_g}{r} \left(\frac{\gamma L_c}{\beta}\right)^2 << 1.
\end{align}
This result also explains the formation of harmonic states in QCLs (a.k.a. Turing rolls). Because the frequency cannot be stably swept by more than the gain bandwidth during a round-trip period, when the dispersion is small or the nonlinearity is large the only stable steady-state solutions to the NLSE will be Turing patterns. These are FM profiles that have N sweeps per round trip, and therefore only sweep over 1/Nth the bandwidth.

Finally, we discuss the role of the Kerr nonlinearity. It was previously shown numerically in \cite{opacakTheoryFrequencyModulatedCombs2019} that the Kerr nonlinearity can shift the dispersion range over which combs form. The mean-field simulations predicted this as well, but at first glance it is unclear why this should be the case. After all, for a constant envelope the Kerr nonlinearity should merely provide a carrier-envelope phase given by Equation (\ref{kerrceo}). This marginally changes the frequency, but does not change the phase profile. The solution to this conundrum lies in the interplay of gain curvature and Kerr nonlinearity. As we showed above, gain curvature reduces the amplitude of the FM comb away from z=0. For a parabolic amplitude perturbation the Kerr effect can be approximated as inducing $\frac{\partial F}{\partial t}=-i \gamma_K' |A|^2 F$, and a dip in the intracavity power will act to induce a chirp on the phase. The phase evolution of the NLSE (Equation \ref{phaseevolve}) becomes
\begin{align}
\frac{\partial \phi}{\partial t} = -\frac{\beta}{2} \left(\frac{\partial \phi}{\partial z}\right)^2 + \gamma |A|^2  (\phi-\langle\phi\rangle) +\frac{1}{r}D_g \gamma_K' \left(\frac{\partial \phi}{\partial z}\right)^2.
\end{align}
In other words, gain curvature and Kerr nonlinearity act together to create an effective dispersion of
\begin{align} \label{deff}
\beta_\textrm{eff} = \beta - \frac{2}{r}D_g \gamma_K'.
\end{align}
Both parameters will shift the valid dispersion range. This is confirmed numerically in Fig. \ref{fig:kerrcurvature}. Fig. \ref{fig:kerrcurvature}a shows the mean field simulation for a laser with a GVD of -800 fs$^2$/mm and no Kerr nonlinearity. With this dispersion value the comb would be hopelessly unstable, since the intensity sag would cause the intensity to dip below zero. However, the addition of a Kerr nonlinearity of $\gamma_K =3.5\times10^{-11}$ m/V$^2$ reduces the dispersion and allows for stable comb production. The analytic theory agrees well with this result, and in particular the chirp calculated numerically agrees with the value expected analytically.

Physically, this process can be thought of within the context of intensity solitons. The effect of gain curvature is to create a dip in the intensity. If the Kerr nonlinearity is allowed to balance the dispersion, this dip can experience gain and will grow into a dark soliton. But this state will exist only transiently, and will eventually cause the intensity to reach zero and destabilize the comb. In other words, the conditions for FM combs to form are precisely the opposite of the conditions for AM combs to form---one must ensure that dispersion and nonlinearity are somewhat \textit{imbalanced} rather than balanced.

\section{Conclusion}

We have shown for the first time that light can obey a nonlinear Schr\"{o}dinger equations whose potential is proportional to its phase. The fundamental solution to this equation is a type of soliton whose frequency is modulated linearly in time, and explains the numerous experimental observations of FM combs in various laser systems. We arrived at this result by deriving a mean-field theory for lasers with cross-steepening, and our results explain the previously-mysterious dynamics of these combs. Our results will pave the way for the development of new types of light sources utilizing these concepts, as the mean field theory derived here is general for many lasers systems.

\appendix
\section{Master equation derivation}\label{sec:meqn}
Our approach uses a master equation formalism and assumes that the laser is described by a two-level system. We assume that the electric field inside the cavity is described by a slowly-varying envelope in both directions, such that
\begin{align}
E_\textrm{full}=E_+(z,t)e^{i(\omega_0 t-k_0 z)}+E_-(z,t)e^{i(\omega_0 t+k_0 z)}+c.c.
\end{align}
Within this formalism, the waveguide losses and dispersion can be found using standard master equations \cite{hausModelockingLasers2000} as
\begin{align} \label{pnl}
\frac{n}{c} \frac{\partial E_\pm}{\partial t}\pm \frac{\partial E_\pm}{\partial z} = &-\frac{\alpha_w}{2} E_\pm +i\frac{1}{2}k''\frac{\partial^2 E_\pm}{\partial t^2} \nonumber \\
&+i \frac{1}{2\omega_0 n \epsilon_0 c} \frac{\partial^2 P_{NL}}{\partial t^2} e^{-i(\omega_0 t\mp k_0 z)}
\end{align}
where n is the refractive index, $k''$ is the dispersion, and $P_{NL}$ is the nonlinear polarization for the system under consideration. Our primary task is to find the last term, which we will call $f_{NL}$.

To find the nonlinear polarization, the gain and cross-steepening are both derived using a standard Maxwell-Bloch formalism \cite{kaertnerUltrafastOptics2010}. If w is used to represent the population inversion and d the coherences, then the dynamics of the system at a particular point in space will be governed by
\begin{align}
\frac{\partial}{\partial t} w &= \frac{2}{i\hbar} \left(d \mu^\ast a^\ast e^{-i\omega_0 t}-d^\ast \mu a e^{i\omega_0 t}\right)-\frac{w-w_0}{T_1} \\
\frac{\partial}{\partial t} d &=\frac{1}{i\hbar} w \mu a e^{i\omega_0 t} + \left(i\omega_0-\frac{1}{T_2}\right)d
\end{align}
where $a\equiv E_+e^{-ik_0 z}+E_-e^{i k_0 z}$ is used to represent both envelopes, $T_1$ and $T_2$ are the respective population and coherence relaxation rates, $w_0$ is the equilibrium population inversion, $\mu$ is the dipole matrix element. Note that we have also implicitly assumed that the central frequency $\omega_0$ is also the energy difference of the system. The coherence can be solved exactly as
\begin{align} \label{dexpr}
d&=e^{i\omega_0 t} \frac{w_0 T_2 \mu}{i \hbar}\left(1+T_2 \partial_t\right)^{-1} a \nonumber\\
&\times \Big[ 1+\frac{2T_1 T_2}{\hbar^2}|\mu|^2 \left(1+T_1 \partial_t\right)^{-1} \nonumber \\ &\hspace{0.25cm}\times\left(a^\ast \left(1+T_2 \partial_t\right)^{-1} a + a\left(1+T_2 \partial_t\right)^{-1}a^\ast\right)\Big]^{-1}
\end{align}
While exact, this expression is somewhat unwieldly due to the presence of the derivatives. Note that if a is constant in time, the term in brackets reduces to the familiar $\left(1+\frac{4T_1 T_2}{\hbar^2}|\mu|^2 |a|^2\right)^{-1}$ intensity saturation function, so we define the saturation intensity $P_s \equiv \left(\frac{4T_1 T_2}{\hbar^2}|\mu|^2 \right)^{-1}$. The nonlinear polarization is found in terms of the atomic density $N$ as
\begin{align} \label{pdefn}
P=-N(\mu^\ast d+\mu d^\ast).
\end{align}
By combining (\ref{pnl}), (\ref{dexpr}), and (\ref{pdefn}), we can find that the positive frequency component of $f_{NL}$ is
\begin{align} \label{fNL}
f_{NL}^{(g)} &=\frac{g_0}{2}\left(1+T_2 \partial_t\right)^{-1} a \Big[ 1+\frac{1}{2P_s} \left(1+T_1 \partial_t\right)^{-1} \nonumber \\ &\times\left(a^\ast \left(1+T_2 \partial_t\right)^{-1} a + a\left(1+T_2 \partial_t\right)^{-1}a^\ast\right)\Big]^{-1} e^{\pm i k_0 z}
\end{align}
where the prefactor has units of gain and is referred to as the small signal gain. In terms of the parameters for a QCL, it can be written as 
\begin{align}
\frac{g_0}{2} \equiv \frac{N \mu^2 \omega_0 w_0 T_2}{2 n \epsilon_0 c \hbar} = \frac{e z_0^2 \omega_0 J T_1 T_2}{2n\epsilon_0 c \hbar L_\textrm{mod}}
\end{align}
where $J$ is the pump current in A/cm$^2$, e is the electron charge, $z_0$ is the dipole moment, and $L_{mod}$ is the module length. To simplify (\ref{fNL}) further, we insert the definition of $a(z,t)$ and expand out the derivatives in terms of a power series, keeping only those terms which have $e^{\mp i k_0 z}$ dependence. We keep only terms up to third order in $E$ and up to first-order in $\partial_t$. We also keep the term that is second-order in $\partial_t$ and first order in $E$, as it is needed to describe gain curvature. Doing this, we find
\begin{align} \label{fullNLeqn} 
f_{NL}^{(g)} &= \frac{g_0}{2}\bigg[ E_\pm-\frac{1}{P_s}(\left|E_\pm\right|^2+2\left|E_\mp\right|^2)  E_\pm \nonumber \\ &\hspace{0.75cm}-T_2 \frac{\partial E_\pm}{\partial t} +T_2^2 \frac{\partial^2 E_\pm}{\partial t^2}\nonumber \\ 
&\hspace{-0.1cm}+\frac{1}{P_s}\bigg( (2T_1+3T_2) \frac{\partial E_\mp^\ast}{\partial t} E_\mp +(T_1+\tfrac{5}{2}T_2) E_\mp^\ast \frac{\partial E_\mp}{\partial t} \nonumber \\
&\hspace{0.7cm}+(T_1+\tfrac{5}{2}T_2) E_\mp^\ast E_\mp \frac{\partial}{\partial t} + (T_1+\tfrac{5}{2}T_2) E_\pm^\ast  \frac{\partial E_\pm}{\partial t} \nonumber \\
&\hspace{0.7cm}+(T_1+\tfrac{3}{2}T_2) \frac{\partial E_\pm^\ast}{\partial t} E_\pm   \bigg) E_\pm \bigg].
\end{align}
Some discussion is in order. The first line is responsible for gain saturation, as it reduces gain in the presence of power. The second line is essentially the system's Lorentzian linewidth---the first derivative is responsible for a refractive index shift, while the second derivative is responsible for gain curvature. The final three lines are responsible for cross- and self-steepening. The last three terms only shift group delay by a small amount, and are typically neglected.

Finally, we consider the Kerr nonlinearity. In the presence of a third-order optical nonlinearity the nonlinear polarization can be written as
\begin{align}
	P_{NL}&=\epsilon_0 \chi^{(3)} E_\textrm{full}^3 \nonumber \\
	&=\epsilon_0 \chi^{(3)} 3(|E_\pm|^2+2|E_\mp|^2) E_\pm e^{i(\omega_0 t \mp k_0 z)}
\end{align}
where we have once again expanded the nonlinearity and have kept only the terms that have $e^{i(\omega_0 t \mp k_0 z)}$ dependence. The Kerr term can then be written as
\begin{align}
f_{NL}^{(K)}&= -i \gamma_K (|E_\pm|^2+2|E_\mp|^2) E_\pm 
\end{align}
where $\gamma_K\equiv \frac{3 \chi^{(3)} \omega_0}{2nc}$. Note that the factor of 2 on $|E_\mp|^2$ is needed to correctly describe cross phase modulation.

Putting it all together, we have
\begin{align}
\frac{n}{c} &\frac{\partial E_\pm}{\partial t}\pm \frac{\partial E_\pm}{\partial z} = -\frac{\alpha_w}{2} E_\pm +i\frac{1}{2}k''\frac{\partial^2 E_\pm}{\partial t^2} \nonumber \\
&\hspace{2.4cm}-i \gamma_K (|E_\pm|^2+2|E_\mp|^2) E_\pm \nonumber \\
&+ \frac{g_0}{2}\bigg[ E_\pm-\frac{1}{P_s}(\left|E_\pm\right|^2+2\left|E_\mp\right|^2)  E_\pm \nonumber \\ &\hspace{0.75cm}-T_2 \frac{\partial E_\pm}{\partial t} +T_2^2 \frac{\partial^2 E_\pm}{\partial t^2}\nonumber \\ 
&\hspace{-0.1cm}+\frac{1}{P_s}\bigg( (2T_1+3T_2) \frac{\partial E_\mp^\ast}{\partial t} E_\mp +(T_1+\tfrac{5}{2}T_2) E_\mp^\ast \frac{\partial E_\mp}{\partial t} \nonumber \\
&\hspace{0.7cm}+(T_1+\tfrac{5}{2}T_2) E_\mp^\ast E_\mp \frac{\partial}{\partial t} + (T_1+\tfrac{5}{2}T_2) E_\pm^\ast  \frac{\partial E_\pm}{\partial t} \nonumber \\
&\hspace{0.7cm}+(T_1+\tfrac{3}{2}T_2) \frac{\partial E_\pm^\ast}{\partial t} E_\pm   \bigg) E_\pm \bigg].
\end{align}
For reference, the parameters for the simulations and calculations in this work are included in Table \ref{tab:params}.

\begin{table}
	\begin{ruledtabular}
		\begin{tabular}{lp{15mm}p{15mm}p{20mm}}
			\textrm{Name}&
			\textrm{Symbol}&
			\textrm{Value}\\
			\colrule\\[-0.2cm]
			Population lifetime & $T_1$ & 4 ps \\
			Coherence lifetime & $T_2$ & 50 fs \\
			Refractive index & $n$ & 3.3 \\
			Waveguide losses & $\alpha_w$ & 4 cm$^{-1}$ \\
			Dipole moment & $z_0$ & 2.3 nm\\
			Mirror 1 reflectivity & $R_1$ & 1 \\
			Mirror 2 reflectivity & $R_2$ & 0.09 \\
			Cavity length & $L_c$ & 4 mm \\
			Module length & $L_\textrm{mod}$ & 58 nm \\
			Wavelength & $\lambda_0$ & 8 $\mu$m \\
			Current density & $J$ & 1150\,A/cm$^2$ \\
			Node spacing & $\Delta z$ & 8 $\mu$m \\
			Dispersion & $k''$ & variable \\
			Kerr nonlinearity & $\gamma_K$ & variable \\
		\end{tabular}
	\end{ruledtabular}	\caption{\label{tab:params}%
		Parameters used for all of the simulations and calculations presented here.
	}
\end{table}
\bibliography{zotero_library}

\end{document}